\documentclass[epj]{webofc}
\usepackage[utf8]{inputenc}
\usepackage[varg]{txfonts}   
\usepackage{booktabs}
\usepackage{xcolor}
\definecolor{darkred}{rgb}{0.4,0.0,0.0}
\definecolor{darkgreen}{rgb}{0.0,0.4,0.0}
\definecolor{darkblue}{rgb}{0.0,0.0,0.4}
\usepackage[bookmarks,linktocpage,colorlinks,
    linkcolor = darkred,
    urlcolor  = darkblue,
    citecolor = darkgreen]{hyperref}
\usepackage{subfigure}
\usepackage{slashed}
\usepackage[export]{adjustbox}
\usepackage{dcolumn}
\wocname{EPJ Web of Conferences}
\woctitle{Lattice2017}
\newcommand{\minidiag}[2]{\begin{minipage}{2cm} \includegraphics[height=2cm]{#2} \end{minipage}}

\renewcommand{\O}{\mathcal{O}}
\newcommand{\p}{\partial}

\renewcommand{\Im}{\mathrm{Im}}

\newcommand{\GeV}{\,\text{GeV}}
\newcommand{\<}{\langle}
\renewcommand{\>}{\rangle}

\newcommand{\remark}[1]{}

\newcommand{\mytag}{\\[-\baselineskip] \stepcounter{equation}\tag{\theequation}}

\begin{document}
%
\selectlanguage{english}
\title{%
Hadronic light-by-light contribution to $\mathbf{(g-2)_\mu}$:\\ a dispersive approach
}
\author{%
\firstname{Gilberto} \lastname{Colangelo}\inst{1}\fnsep\thanks{Speaker,
  \email{gilberto@itp.unibe.ch} } ,
\firstname{Martin} \lastname{Hoferichter}\inst{2} ,
\firstname{Massimiliano}  \lastname{Procura}\inst{3} \and
\firstname{Peter}  \lastname{Stoffer}\inst{4}
}
\institute{%
Albert Einstein Center for Fundamental Physics, Institute for Theoretical Physics\\
University of Bern, Sidlerstrasse 5, 3012 Bern, Switzerland 
\and
Institute for Nuclear Theory, University of Washington, Seattle, WA 98195-1550, USA
\and
Fakult\"at f\"ur Physik, Universit\"at Wien, Boltzmanngasse 5, 1090 Wien,
Austria
\and
Department of Physics, University of California at San Diego, La Jolla, CA 92093, USA
}
\abstract{%
After a brief introduction on ongoing experimental and theoretical
activities on $(g-2)_\mu$, we report on recent progress in approaching the
calculation of the hadronic light-by-light contribution with dispersive
methods. General properties of the four-point function of 
the electromagnetic current in QCD, its Lorentz decomposition and
dispersive representation are discussed. On this basis a numerical
estimate for the pion box contribution and its rescattering corrections is
obtained. We conclude with an outlook for this approach to the calculation
of hadronic light-by-light.
}
\maketitle
\section{Introduction}\label{intro}
The measured value of the anomalous magnetic moment of the muon $a_\mu$,
obtained by the BNL E821 experiment~\cite{Bennett:2006fi}, represents a
puzzle for the standard model (SM): it differs by about three standard
deviations from the calculated value (see
e.g.~\cite{Hagiwara:2017zod,Davier:2016iru}). Taken at face value this is a
serious discrepancy, but before claiming a 
real crisis for the SM or plain discovery of new physics, it is important
to make sure that systematic effects, either on the theory or on the
experimental side, have not been underestimated. Experimentally this
requires redoing the measurement, ideally in a completely new setting. This
is the aim of the Muon $g-2$ experiment~\cite{Grange:2015fou} which has
started to run at Fermilab and aims to reduce the final uncertainty reached
by the BNL E821 experiment by about a factor four. This experiment is reusing
the same ring and conceptual design of the Brookhaven experiment. A
completely different setting has instead been adopted by the J-PARC E34
experiment~\cite{Saito:2012zz}, which however still needs to be approved
and will start running only in a few years from now.

On the theory side there has been impressive progress in recent years in
the calculation of pure QED
contributions~\cite{Aoyama:2012fc,Aoyama:2012wk,Aoyama:2014sxa,Laporta:2017okg}. Also
electroweak contributions, which are known to two loops, are under good
control and have survived further more recent tests and double
checks~\cite{Gnendiger:2013pva,Ishikawa:2017ouv}. None of these
improvements or checks has had any significant impact neither on the
central value nor on the error, essentially because their contributions to
the theoretical uncertainty is negligible with respect to that coming from
hadronic contributions. The latter are indeed the most crucial (for the
error estimate) and critical (for the central value) contributions to the
SM value and a lot of theoretical activity has been and still is devoted to
improving their estimate. Quite a substantial part of it concerns lattice
calculations. 

Hadronic contributions can be classified based on the order in $\alpha$ and
the topology. At order $\alpha^2$ there is the hadronic vacuum polarization
contribution to the vertex-correction diagram, usually called {\em tout
  court} hadronic vacuum polarization (HVP). At order $\alpha^3$, there are
next-to-leading (NLO) order HVP diagrams as well as hadronic light-by-light
(HLbL) contributions. At order $\alpha^4$ there are NNLO HVP diagrams and
NLO HLbL diagrams. Going beyond LO for each topology is important for the
central value estimate and for checking that there are no surprises, but
for the uncertainty estimate it is the LO contributions which matter. As it
turns out, the two topologies contribute about the same to the total
uncertainty, even though HLbL is suppressed by one order of $\alpha$. The
reason is simple: the calculation of the HVP is based on an exact
relation, derived from analyticity and unitarity, which allows one to
express this contribution as an integral over the measurable cross section
$\sigma(e^+ e^- \to \mbox{hadrons})$. The latter cross section has actually
been measured with high accuracy, especially at low energy, which is the
region contributing with the highest weight, so that the calculation of the
relevant integral can be performed with subpercent accuracy. For HLbL
such a relation did not exist until recently and has been derived in a
series of recent
papers~\cite{Colangelo:2014dfa,Colangelo:2014pva,Colangelo:2015ama,Colangelo:2017qdm,Colangelo:2017fiz}.
This is not as simple and effective as the one for HVP, where all intermediate
states contribute in the same form: for HLbL different intermediate states
appear in different integrals and the explicit form of these integrals has
been derived only for up to two-particle intermediate states. Moreover, the
measurable quantities that enter these integrals ({\em e.g.} the $\gamma^*
\gamma^* \to \pi \pi$ helicity amplitudes) have not yet been
measured other than in corners of the phase space (for two or one real
photons). Complete estimates of the HLbL which have appeared so far are
therefore based on models~\cite{Bijnens:1995xf,Bijnens:2001cq,Hayakawa:1995ps,Hayakawa:1996ki,Hayakawa:1997rq,Melnikov:2003xd}
(an alternative dispersive approach, where the whole muon form factor
is treated dispersively has been proposed in Ref.~\cite{Pauk:2014rfa}).

In this contribution I will briefly report on recent progress in the
calculation of the HLbL contribution on the lattice and discuss in some
more detail the dispersive approach we have developed, in particular
illustrating our first numerical estimate of the pion box and the
corresponding rescattering contribution. The status of the calculations 
of the HVP contribution is covered by the talk given by Christoph
Lehner~\cite{LehnerLat17}.

\section{Hadronic light-by-light on the lattice}\label{HLbl-lattice}

Two lattice collaborations have started different approaches to calculate
the HLbL contribution. Until recent years it was not known how this
contribution could be calculated on the lattice and at the present stage it
is not clear which of the approaches tried so far will be the most
effective and will reach the highest precision in the long run.

\subsection{The RBC/UKQCD approach}\label{sec:latticeRBC}
The first attempt to perform a lattice calculation of the HLbL contribution
to $(g-2)_\mu$ is due to Blum et al.~\cite{Blum:2014oka}. The method
proposed relied on putting on the lattice not only quarks and gluons, but
also photons and muons and integrating over the two loops involving the
muon and photon propagators with Monte Carlo methods on the lattice. The
calculation is nontrivial, not only because the hadronic object to be
considered is complicated, but also because, by choosing to include also
the photons in the calculation, all the difficulties related to having
massless photons on the lattice affect this calculation too. In particular
one expects quite significant finite-volume effects due to the photons
(see~\cite{IzubuchiLat17}). The 
first calculation only considered connected diagrams and was performed with
unphysical quark masses, whereas a more recent one~\cite{Blum:2016lnc,BlumLat17},
performed along the same lines, included the leading disconnected diagram
and was done at the physical point. No attempt at taking the continuum
limit was made yet, since the calculation was performed at a single lattice
spacing. 

The result obtained reads:
\begin{equation}
a_\mu^\mathrm{HLbL}=(5.35 \pm  1.35) \cdot 10^{-10} \; \;,
\end{equation}
which is about half of the most recent estimates based on
models~\cite{Prades:2009tw,Jegerlehner:2009ry}. Given the exploratory
nature of the calculation and the fact that it has been performed at a
single lattice spacing, it is premature to discuss about numerical 
differences. The main message to be taken from the impressive RBC/UKQCD 
effort is that the calculation is possible with current computers and that 
the chosen approach seems to work well. Further efforts in this direction
promise to provide a number with a controlled uncertainty estimate, however
the problems related to the finite-volume effects can be more efficiently
solved with a different approach, proposed by the Mainz group (see below),
which has recently been adopted also by the RBC/UKQCD
collaboration~\cite{Blum:2017cer}. 

\subsection{The Mainz approach}\label{sec:latticeMainz}
The approach followed by the Mainz group is based on the following formula
which expresses the contribution as an integral in position (rather
than in momentum) space~\cite{Asmussen:2016lse}:
\begin{equation}
a_\mu^{\rm HLbL} = \frac{m e^6}{3}  \int d^4y
   \Big[
      \int d^4x
      \underbrace{\bar{\cal L}_{[\rho,\sigma];\mu\nu\lambda}(x,y)}_{\rm QED}\;
      \underbrace{i\widehat\Pi_{\rho;\mu\nu\lambda\sigma}(x,y)}_{\rm QCD}
   \Big] \; \;.
\end{equation}
The advantage of such an approach is that the photon propagators are
handled analytically and taken care of by the kernel $\bar{\cal
  L}_{[\rho,\sigma];\mu\nu\lambda}(x,y)$, which has been calculated
exactly. In this way all the problems related to the formulation of QED on 
the lattice, and in particular in finite volume, are completely
overcome. Effectively one is using QED in infinite volume. 
Moreover it is much closer to what is actually calculated on the lattice,
where it is position space which is discretized. 
The approach has been successfully tested numerically by inserting in
$i\widehat\Pi_{\rho;\mu\nu\lambda\sigma}(x,y)$ the explicit expression for
a muon loop and performing the calculation on the lattice. The outcome
reproduced to percent accuracy the known result for the muonic
light-by-light contribution~\cite{AsmussenLat17}. A similar test has also
been successfully performed by the RBC/UKQCD
collaboration~\cite{Blum:2017cer}. 

The Mainz group has also followed a different path, namely to calculate
explictly the pion transition form factor on the lattice. This
is only one of the contributions to the HLbL tensor (see below for a
precise definition thereof), but arguably the most important one. Moreover,
experimental data on this form factor are available only for the
singly-virtual case, and efforts to measure the doubly-virtual
configurations are plagued by serious difficulties. On the other hand this
is not a particularly difficult calculation on the lattice, and the one
done by the Mainz group shows that it can be performed with a very good
accuracy:
\begin{equation}
a_{\mu,LMD+V}^\mathrm{HLbL,\pi^0}= (6.50 \pm 0.83) \cdot 10^{-10}
\end{equation}
of about 13\%~\cite{Gerardin:2016cqj,GerardinLat17}: investing more efforts and
more computer time in such a calculation, has the potential to bring the
accuracy of this contribution to well below 10\%, which is sufficient for
the present purpose. Note that although the calculation is done from first
principles and aims to be model independent, in order to perform the full
integral over the photon and muon loops one needs to model the $q_i^2$
dependence of the pion transition form factor even beyond the region where
the lattice calculation has been made. Technically this means that one
chooses a parametrization to describe the $q^2$ dependence and fixes the
values of the parameters to fit the lattice data. In this way the result of
the integration depends on the chosen parametrization --- this explains the
label ``LMD+V'' attached to the result, which indicates the kind of
parametrization used. The final aim is to cover on the lattice a large enough
region in $q^2$ such that the model-dependence due to the parametrization
becomes negligible.

\section{Dispersive approach: preliminaries}\label{prelim}
In order to attack the calculation of the HLbL dispersively a few
preliminary steps are necessary. For simpler objects, like the two-point
function of the electromagnetic current, which is relevant for the HVP
contribution, these steps are usually performed without even mentioning
them, because they are almost trivial. For the two-point function of the
electromagnetic current Lorentz invariance allows one to decompose the
tensor into two independent structures. Gauge invariance reduces the two
independent structures to a single one. Going from the two- to the
four-point function the increase in complexity is baffling: the number of
independent Lorentz structures jumps from 2 to 138 (136 in 4 dimensions,
see~\cite{Eichmann:2015nra}). Also the implementation of gauge invariance
becomes significantly more complex -- especially if one wants to obtain a
basis which is free from kinematic singularities and zeros -- but luckily a
general procedure has been devised long ago by Bardeen and
Tung~\cite{Bardeen:1969aw}, with an important addendum pointed out by
Tarrach~\cite{Tarrach:1975tu}. This can be applied also in this case
without special difficulties other than those due to the inherent
complexity of the problem. 

\subsection{Lorentz and gauge invariant decomposition}\label{BTT}

\begin{figure}[t]
	\centering
	\includegraphics[width=5.5cm]{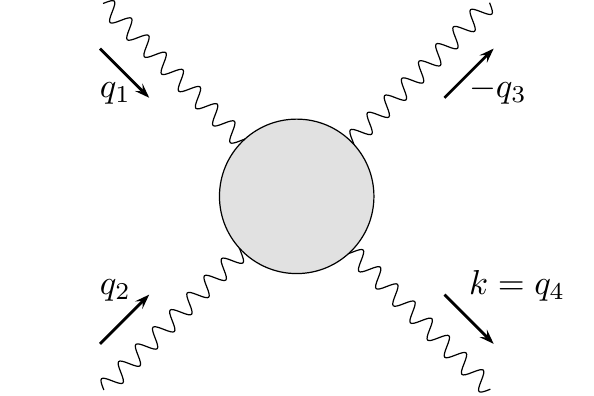}
	\caption{Kinematics of the light-by-light scattering amplitude.}
	\label{img:FullHLbL}
\end{figure}

The HLbL tensor is the Green's function of four electromagnetic
currents, evaluated in pure QCD: 
\begin{align}
	\label{eq:HLbLTensorDefinition}
	\Pi^{\mu\nu\lambda\sigma}(q_1,q_2,q_3) = -i \int d^4x \, d^4y \, d^4z \, e^{-i(q_1 \cdot x + q_2 \cdot y + q_3 \cdot z)} \< 0 | T \{ j_\mathrm{em}^\mu(x) j_\mathrm{em}^\nu(y) j_\mathrm{em}^\lambda(z) j_\mathrm{em}^\sigma(0) \} | 0 \> .
\end{align}
The electromagnetic current above is built out of the three lightest quarks
only:
\begin{align}
	j_\mathrm{em}^\mu := \bar q Q \gamma^\mu q ,
\end{align}
where $q = ( u , d, s )^T$ and $Q = \mathrm{diag}(\frac{2}{3}, -\frac{1}{3}, -\frac{1}{3})$.
We define
\begin{align}
	q_4 := k = q_1 + q_2 + q_3 ,
\end{align}
and illustrate the kinematics in Fig.~\ref{img:FullHLbL}.

As invariant variable we adopt the usual Mandelstam variables:
\begin{align}
	s := (q_1+q_2)^2, \quad t := (q_1+q_3)^2, \quad u := (q_2 + q_3)^2,
        \qquad s + t + u = \sum_{i=1}^4 q_i^2 =: \Sigma
\end{align}
(the limit $k^2 = 0$ will be considered later).
The Ward--Takahashi identities implied by gauge invariance have the form
\begin{align}
	\label{eq:WardIdentitiesHLbLTensor}
	\{q_1^\mu, q_2^\nu, q_3^\lambda, q_4^\sigma\} \Pi_{\mu\nu\lambda\sigma}(q_1,q_2,q_3) = 0 .
\end{align}

\subsection{Tensor decomposition}

\label{sec:HLbLTensorBTTDecomposition}

As mentioned above the HLbL tensor can be decomposed into 138 Lorentz
structures~\cite{Karplus:1950zza, Leo:1975fb, Bijnens:1996wm}: 
\begin{align}
	\begin{split}
		\label{eq:HLbLTensor138StructuresLSM}
		\Pi^{\mu\nu\lambda\sigma} &= g^{\mu\nu} g^{\lambda\sigma} \, \Pi^1 + g^{\mu\lambda} g^{\nu\sigma} \, \Pi^2 + g^{\mu\sigma} g^{\nu\lambda} \, \Pi^3 
			+ \sum_{\substack{i=2,3,4 \\ j=1,3,4}}
                        \sum_{\substack{k=1,2,4 \\ l=1,2,3}} q_i^\mu
                        q_j^\nu q_k^\lambda q_l^\sigma \, \Pi^4_{ijkl}  \\
			&  +
                        \sum_{\substack{i=2,3,4 \\ j=1,3,4}}
                        g^{\lambda\sigma} q_i^\mu q_j^\nu \, \Pi^5_{ij} + \sum_{\substack{i=2,3,4 \\ k=1,2,4}} g^{\nu\sigma} q_i^\mu q_k^\lambda \, \Pi^6_{ik} + \sum_{\substack{i=2,3,4 \\ l=1,2,3}} g^{\nu\lambda} q_i^\mu q_l^\sigma \, \Pi^7_{il}  \\
			& + \sum_{\substack{j=1,3,4 \\ k=1,2,4}}
                        g^{\mu\sigma} q_j^\nu q_k^\lambda \, \Pi^8_{jk} +
                        \sum_{\substack{j=1,3,4 \\ l=1,2,3}} g^{\mu\lambda}
                        q_j^\nu q_l^\sigma \, \Pi^9_{jl} + \sum_{\substack{k=1,2,4 \\ l=1,2,3}} g^{\mu\nu}
                        q_k^\lambda q_l^\sigma \, \Pi^{10}_{kl} \qquad
                        \qquad . 
	\end{split}
\end{align}
The 138 scalar functions $\{ \Pi^1, \Pi^2, \Pi^3, \Pi^4_{ijkl}, \Pi^5_{ij},
\Pi^6_{ik}, \Pi^7_{il}, \Pi^8_{jk}, \Pi^9_{jl}, \Pi^{10}_{kl}\} $
depend on six independent kinematic variables: the two Mandelstam variables
$s$ and $t$ and the virtualities $q_1^2$, $q_2^2$, $q_3^2$, and
$q_4^2$. They are free of kinematic singularities but since they have to
fulfill kinematic constraints required by gauge invariance, they must have
kinematic zeros. The Ward identities~\eqref{eq:WardIdentitiesHLbLTensor} impose
95 linearly independent relations on the scalar functions, reducing the set
to 43 functions. To obtain these we apply the recipe devised by Bardeen and
Tung~\cite{Bardeen:1969aw}, but this does not lead to a minimal
basis free of kinematic singularities, as shown by
Tarrach~\cite{Tarrach:1975tu}. Following the latter we have constructed a
redundant set of 54 structures, which is free of kinematic 
singularities and zeros.

The resulting representation of the HLbL tensor which we have obtained in
this way reads
\begin{align}
	\label{eqn:HLbLTensorKinematicFreeStructures}
	\Pi^{\mu\nu\lambda\sigma} &= \sum_{i=1}^{54} T_i^{\mu\nu\lambda\sigma} \Pi_i , 
\end{align}
where
\begin{align}
		\label{eq:HLbLBTTStructures}
		T_1^{\mu\nu\lambda\sigma} &= \epsilon^{\mu\nu\alpha\beta}
                \epsilon^{\lambda\sigma\gamma\delta} {q_1}_\alpha
                        {q_2}_\beta {q_3}_\gamma {q_4}_\delta , \nonumber \\
		T_4^{\mu\nu\lambda\sigma} &= \Big(q_2^\mu q_1^\nu - q_1
                \cdot q_2 g^{\mu \nu} \Big) \Big( q_4^\lambda q_3^\sigma -
                q_3 \cdot q_4 g^{\lambda \sigma} \Big) , \nonumber \\
		T_7^{\mu\nu\lambda\sigma} &= \Big(q_2^\mu q_1^\nu - q_1
                \cdot q_2 g^{\mu \nu } \Big) \Big( q_1 \cdot q_4
                \left(q_1^\lambda q_3^\sigma -q_1 \cdot q_3 g^{\lambda
                  \sigma} \right) + q_4^\lambda q_1^\sigma q_1 \cdot q_3 -
                q_1^\lambda q_1^\sigma q_3 \cdot q_4 \Big) , \nonumber \\
		 T_{19}^{\mu\nu\lambda\sigma} &= \Big( q_2^\mu q_1^\nu -
                 q_1 \cdot q_2 g^{\mu \nu } \Big) \Big(q_2 \cdot q_4
                 \left(q_1^\lambda q_3^\sigma - q_1 \cdot q_3
                 g^{\lambda\sigma} \right)+q_4^\lambda q_2^\sigma q_1 \cdot
                 q_3 - q_1^\lambda q_2^\sigma q_3 \cdot q_4 \Big) ,
                 \nonumber \\
		T_{31}^{\mu\nu\lambda\sigma} &= \Big(q_2^\mu q_1^\nu -
                q_1\cdot q_2 g^{\mu\nu}\Big) \Big(q_2^\lambda q_1\cdot q_3
                - q_1^\lambda q_2\cdot q_3\Big) \Big(q_2^\sigma q_1\cdot
                q_4 - q_1^\sigma q_2\cdot q_4\Big) , \nonumber \\
		T_{37}^{\mu\nu\lambda\sigma} &= \Big( q_3^\mu q_1\cdot q_4 - q_4^\mu q_1\cdot q_3\Big) \begin{aligned}[t]
			& \Big( q_3^\nu q_4^\lambda q_2^\sigma - q_4^\nu q_2^\lambda q_3^\sigma + g^{\lambda\sigma} \left(q_4^\nu q_2\cdot q_3 - q_3^\nu q_2\cdot q_4\right) \\
			& + g^{\nu\sigma} \left( q_2^\lambda q_3\cdot q_4 -
                  q_4^\lambda q_2\cdot q_3 \right) + g^{\lambda\nu} \left(
                  q_3^\sigma q_2\cdot q_4 - q_2^\sigma q_3\cdot q_4 \right)
                  \Big) , \end{aligned} \nonumber \\
		T_{49}^{\mu\nu\lambda\sigma} &= q_3^\sigma  \begin{aligned}[t]
				& \Big( q_1\cdot q_3 q_2\cdot q_4 q_4^\mu
                  g^{\lambda\nu} - q_2\cdot q_3 q_1\cdot q_4 q_4^\nu
                  g^{\lambda\mu} + q_4^\mu q_4^\nu \left( q_1^\lambda
                  q_2\cdot q_3 - q_2^\lambda q_1\cdot q_3 \right) \nonumber
                  \\
				& + q_1\cdot q_4 q_3^\mu q_4^\nu
                  q_2^\lambda - q_2\cdot q_4 q_4^\mu q_3^\nu q_1^\lambda +
                  q_1\cdot q_4 q_2\cdot q_4 \left(q_3^\nu g^{\lambda\mu} -
                  q_3^\mu g^{\lambda\nu}\right) \Big) \end{aligned}
                \\
			& - q_4^\lambda \begin{aligned}[t]
				& \Big( q_1\cdot q_4 q_2\cdot q_3 q_3^\mu
                  g^{\nu\sigma} - q_2\cdot q_4 q_1\cdot q_3 q_3^\nu
                  g^{\mu\sigma} + q_3^\mu q_3^\nu \left(q_1^\sigma q_2\cdot
                  q_4 - q_2^\sigma q_1\cdot q_4\right)  \\
				& + q_1\cdot q_3 q_4^\mu q_3^\nu q_2^\sigma
                  - q_2\cdot q_3 q_3^\mu q_4^\nu q_1^\sigma + q_1\cdot q_3
                  q_2\cdot q_3 \left( q_4^\nu g^{\mu\sigma} - q_4^\mu
                  g^{\nu\sigma} \right) \Big) \end{aligned}  \\
			& + q_3\cdot q_4 \Big(\left(q_1^\lambda q_4^\mu -
                q_1\cdot q_4 g^{\lambda\mu}\right) \left(q_3^\nu q_2^\sigma
                - q_2\cdot q_3 g^{\nu\sigma}\right) - \left(q_2^\lambda
                q_4^\nu - q_2\cdot q_4 g^{\lambda\nu}\right) \left(q_3^\mu
                q_1^\sigma - q_1\cdot q_3 g^{\mu\sigma}\right)\Big) . \quad
                \nonumber 
\end{align}

All the remaining structures are just crossed versions of the above seven
structures, as shown in Ref.~\cite{Colangelo:2015ama}.  Since the HLbL
tensor $\Pi^{\mu\nu\lambda\sigma}$ is totally crossing symmetric, the
scalar functions $\Pi_i$ have to fulfill exactly the same crossing
properties of the corresponding Lorentz structures, which we have not given
here, but which can again be found in~\cite{Colangelo:2015ama}.  Therefore,
only seven different scalar functions $\Pi_i$ appear, together with their
crossed versions. These 54 scalar functions are free of kinematic
singularities and zeros and hence fulfill a Mandelstam
representation. They are suitable quantities for a dispersive description.

\subsection{Master formula}\label{MF}

As is well known, using projector techniques and angular averaging
(see \cite{Knecht:2001qf,Jegerlehner:2008zza}), the anomalous magnetic moment of
the muon can be expressed as 
\begin{align}
	\begin{split}
		a_\mu &= \mathrm{Tr}\left( \left(\frac{1}{12} \gamma^\mu - \frac{1}{3} \left( \frac{p^\mu \slashed p}{m_\mu^2} \right) - \frac{1}{4} \frac{p^\mu}{m_\mu} \right) V_\mu(p) \right) - \frac{1}{48 m_\mu} \mathrm{Tr}\left( (\slashed p + m_\mu) [\gamma^\mu,\gamma^\rho] (\slashed p + m_\mu) \Gamma_{\mu\rho}(p) \right) ,
	\end{split}
\end{align}
where now $p^2 = m_\mu^2$.\footnote{Note that $k$ is defined as outgoing,
  resulting in the different sign of the second term with respect to
  \cite{Jegerlehner:2008zza}.} 

The contribution of the HLbL tensor to $a_\mu$, represented
diagrammatically as
\begin{align}
	\minidiag{HLbL}{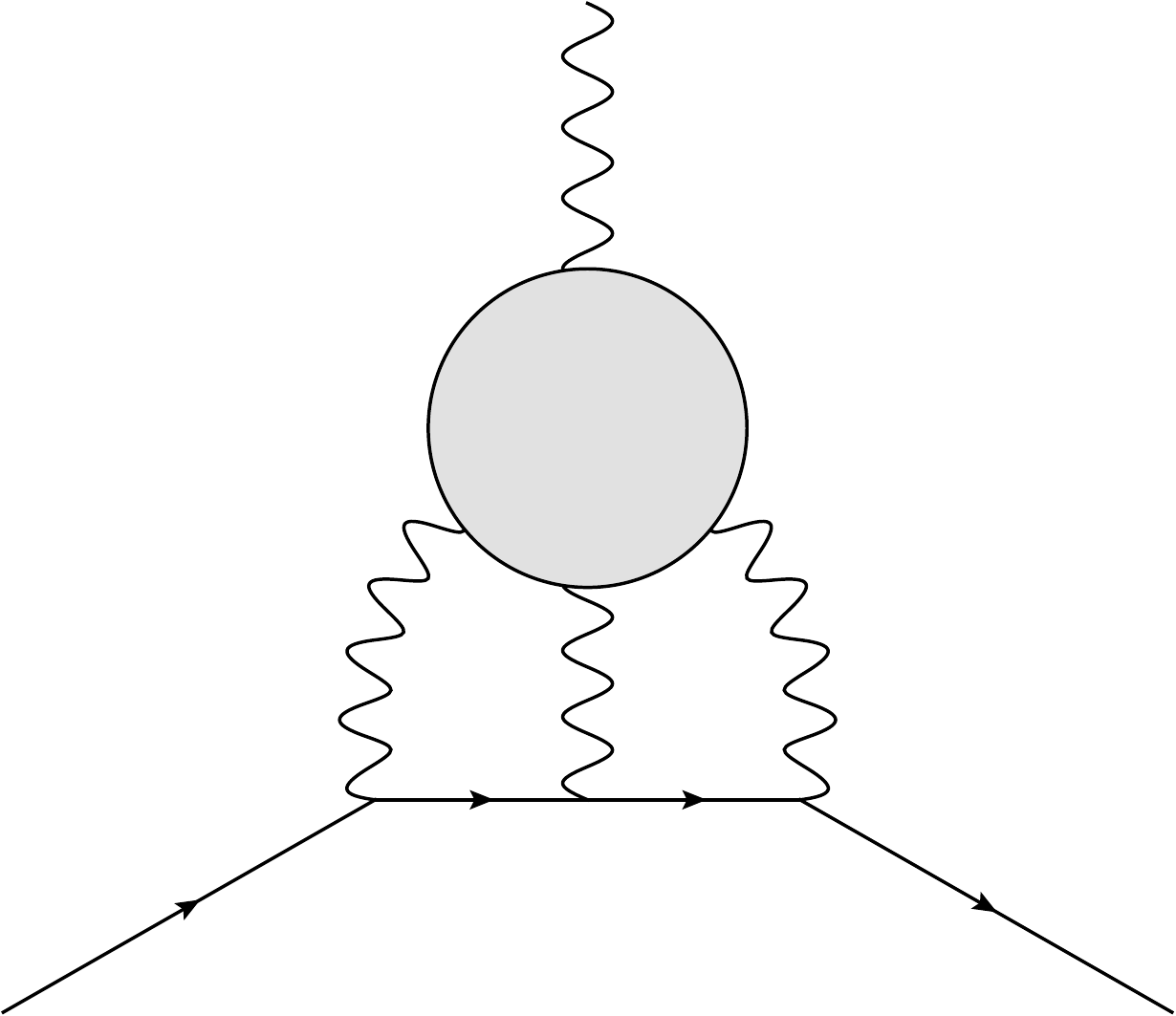} = (-ie) \bar u(p_2) \Gamma^\mu_\mathrm{HLbL}(p_1,p_2) u(p_1) ,
\end{align}
can be written as
\begin{align}
	\begin{split}
		\Gamma^\sigma_\mathrm{HLbL}(p_1,p_2) &= - e^6 \int \frac{d^4q_1}{(2\pi)^4} \frac{d^4q_2}{(2\pi)^4} \gamma_\mu \frac{(\slashed p_2 + \slashed q_1 + m_\mu)}{(p_2+q_1)^2 - m_\mu^2} \gamma_\lambda  \frac{(\slashed p_1 - \slashed q_2 + m_\mu)}{(p_1-q_2)^2 - m_\mu^2} \gamma_\nu \\
			& \quad \times \frac{1}{q_1^2 q_2^2 (p_1-p_2-q_1-q_2)^2} \Pi^{\mu\nu\lambda\sigma}(q_1,q_2,p_1-p_2-q_1-q_2) .
	\end{split}
\end{align}
The HLbL tensor has been defined
in~\eqref{eq:HLbLTensorDefinition}. Differentiating the fourth Ward
identity in~\eqref{eq:WardIdentitiesHLbLTensor} with respect to
$k_\rho=(q_1+q_2+q_3)_\rho$ yields 
\begin{align}
	\label{eq:DiffWardIdentity}
	\Pi_{\mu\nu\lambda\rho}(q_1,q_2,k-q_1-q_2) = - k^\sigma \frac{\p}{\p k^\rho} \Pi_{\mu\nu\lambda\sigma}(q_1,q_2,k-q_1-q_2) .
\end{align}
It was already argued in~\cite{Aldins:1970id} that $\Pi_{\mu\nu\lambda\sigma}$
vanishes linearly with $k$ ({\em i.e.}\ the derivative contains no singularity),
and so must $\Gamma_\sigma^\mathrm{HLbL}$. This is easily verified with our
tensor
decomposition~\eqref{eqn:HLbLTensorKinematicFreeStructures}. Therefore, the
HLbL contribution to the anomalous magnetic moment is given by 
\begin{align}
	\label{eq:amuTraceFormula}
	a_\mu^\mathrm{HLbL} = - \frac{1}{48 m_\mu} \mathrm{Tr}\left( (\slashed p + m_\mu) [\gamma^\rho,\gamma^\sigma] (\slashed p + m_\mu) \Gamma_{\rho\sigma}^\mathrm{HLbL}(p) \right) ,
\end{align}
where
\begin{align}
	\Gamma^\mathrm{HLbL}_{\rho\sigma}(p) = \frac{\p}{\p k^\sigma} \Gamma^\mathrm{HLbL}_\rho(p_1,p_2) \bigg|_{k=0} .
\end{align}
We use the Ward identity~\eqref{eq:DiffWardIdentity} to write
\begin{align}
	\begin{split}
		\label{eq:GammaHLbLTwoLoop}
		\Gamma_{\rho\sigma}^\mathrm{HLbL}(p) &= e^6 \int \frac{d^4q_1}{(2\pi)^4} \frac{d^4q_2}{(2\pi)^4} \gamma^\mu \frac{(\slashed p + \slashed q_1 + m_\mu)}{(p+q_1)^2 - m_\mu^2} \gamma^\lambda  \frac{(\slashed p - \slashed q_2 + m_\mu)}{(p-q_2)^2 - m_\mu^2} \gamma^\nu \\
			& \quad \times \frac{1}{q_1^2 q_2^2 (q_1+q_2)^2}
                \frac{\p}{\p k^\rho}
                \Pi_{\mu\nu\lambda\sigma}(q_1,q_2,k-q_1-q_2) \bigg|_{k=0}
                \; ,
	\end{split}
\end{align}
after taking the derivative and the limit $k_\mu \to 0$.

After a number of intermediate steps, which include performing five of the
eight loop integrals by changing to spherical coordinates in four
dimensions and applying Gegenbauer polynomial techniques we have obtained a
master formula for the HLbL contribution to the anomalous magnetic moment
of the muon: 
\begin{align}
	\label{eq:MasterFormula3Dim}
	a_\mu^\mathrm{HLbL} &= \frac{2 \alpha^3}{3 \pi^2} \int_0^\infty dQ_1 \int_0^\infty dQ_2 \int_{-1}^1 d\tau \sqrt{1-\tau^2} Q_1^3 Q_2^3 \sum_{i=1}^{12} T_i(Q_1,Q_2,\tau) \bar \Pi_i(Q_1,Q_2,\tau) ,
\end{align}
where $Q_1 := |Q_1|$, $Q_2 := |Q_2|$. The hadronic scalar functions
$\bar \Pi_i$ are linear combinations of the $\Pi_i$. 
They have to be evaluated for the reduced kinematics
\begin{align}
	\begin{split}
		s &= - Q_3^2 = -Q_1^2 - 2 Q_1 Q_2 \tau - Q_2^2 , \quad t = -Q_2^2 , \quad u = -Q_1^2 , \\
		q_1^2 &= -Q_1^2, \quad q_2^2 = -Q_2^2, \quad q_3^2 = - Q_3^2 = - Q_1^2 - 2 Q_1 Q_2 \tau - Q_2^2 , \quad k^2 = q_4^2 = 0.
	\end{split}
\end{align}
The integral kernels $T_i$, provided in~\cite{Colangelo:2017fiz} are fully
general for any light-by-light process, while the scalar functions $\Pi_i$
parametrize the hadronic content of the master formula. 
In particular, \eqref{eq:MasterFormula3Dim} can be considered a
generalization of the three-dimensional integral formula for the pion-pole
contribution~\cite{Jegerlehner:2009ry}. It is valid for the whole HLbL
contribution and completely generic, {\em i.e.}\ it can be used to compute the
HLbL contribution to $(g-2)_\mu$ for any representation of the HLbL tensor,
{\em i.e.} of the scalar functions. 

Like in the case of the pion-pole contribution~\cite{Knecht:2001qf}, the
master formula~\eqref{eq:MasterFormula3Dim} offers the great advantage of
providing a representation of the HLbL contribution to the $(g-2)_\mu$ in
terms of a three-dimensional integral, which is well-suited for
a direct numerical implementation. In particular, the energy regions
generating the bulk of the contribution can be identified by numerically
integrating over $\tau$ and plotting the integrand as a function of $Q_1$
and $Q_2$~\cite{Knecht:2001qf,Bijnens:2001cq,Abyaneh:2012ak,Pauk:2014rta}.

\subsection{An ordering principle}\label{Order}
An important difference between the two-point function which is relevant
for HVP and the four-point function of HLbL is that the dispersion relation
for the former is in only one variable and that the discontinuity is given
by the imaginary part which, thanks to unitarity, is related to an
observable: the cross section $e^+ e^- \to \mbox{hadrons}$. Many
intermediate states contribute to the discontinuity, but they all
do with the same weight function inside the integral:
\begin{equation}
a_\mu^\mathrm{HVP} = \left(\frac{\alpha m_\mu}{3 \pi}\right)^2
\int_{s_\mathrm{thr}}^\infty \frac{ds}{s^2} \hat{K}(s) R_\mathrm{had}(s)
\quad \mbox{where}\quad 
R_\mathrm{had}(s) = \frac{\sigma(e^+ e^- \to \mbox{hadrons})}{ 4 \pi \alpha(s)^2/3s}
\end{equation}
and $\hat K(s)$ is the integration kernel which grows monotonically from
about $0.63$ at the two-pion threshold up to $1$ at $s=\infty$. The lower
limit of integration $s_\mathrm{thr}$ is equal to $4 M_\pi^2$ at
$\O(\alpha^0)$, but if one includes photons in the hadronic final state
then it becomes $M_{\pi^0}^2$.

In the case of HLbL there is no such a simple formula, first of all because
the independent kinematic variables are two instead of only one (we
consider a dispersion relation at fixed $q_i^2$, so it is only two of the
three Mandelstam variables which are independent). The situation is
analogous to that of a scattering amplitude, which has been studied in
depth and treated with different kinds of dispersion relations. The most
general thing one can do in this case is to write down a Mandelstam
representation (assuming that it holds). This has indeed be done
in~\cite{Colangelo:2015ama}, but then the practical usefulness of such a
representation is limited by the fact that the double spectral functions,
which completely determine the scalar BTT functions, are not observables
and cannot be directly measured in an inclusive manner, such that all
intermediate states contributing are taken into account at the same
time. The only possible way to make use of such a representation is to
consider individual intermediate states and for each of these construct a
relation between the double-spectral function and the relevant
observable. By doing this one obtains a dispersive representation of the
HLbL tensor as a sum of contributions of different, fully specified
intermediate states, and for each term in the sum there is an explicit
relation to the relevant observable.

\begin{figure}[t]
	\centering
	\begin{align*}
		\includegraphics[width=2.5cm,valign=c]{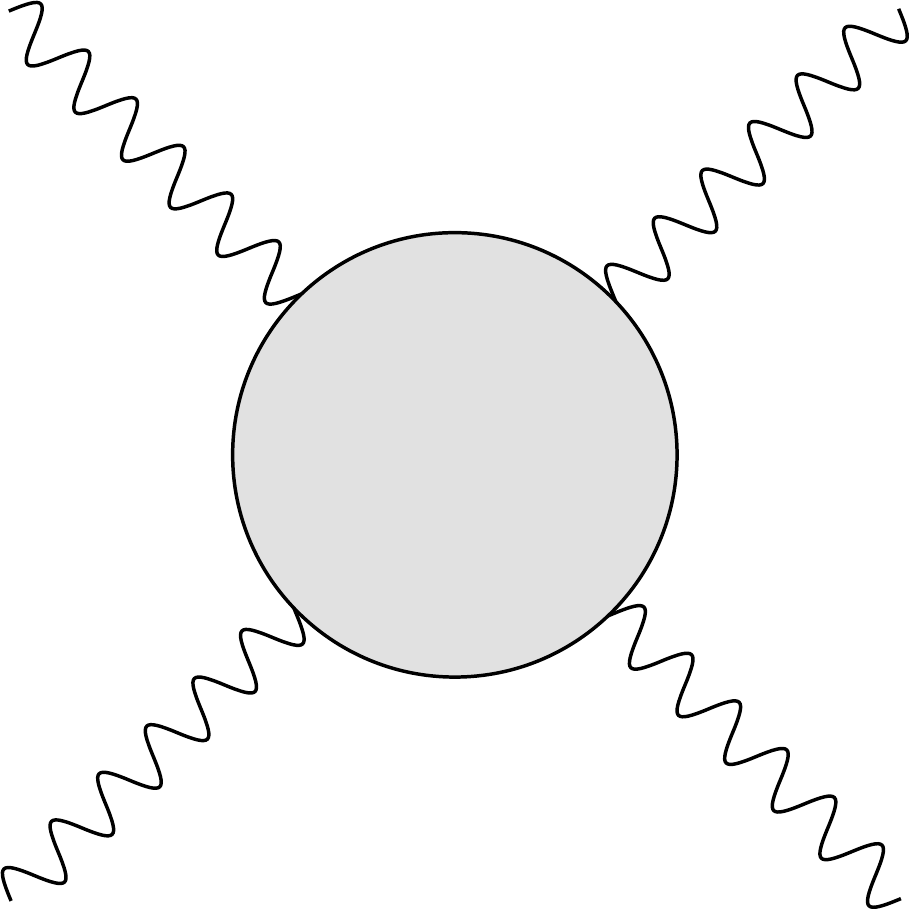}
		 \quad = \quad
		\includegraphics[width=2.5cm,valign=c]{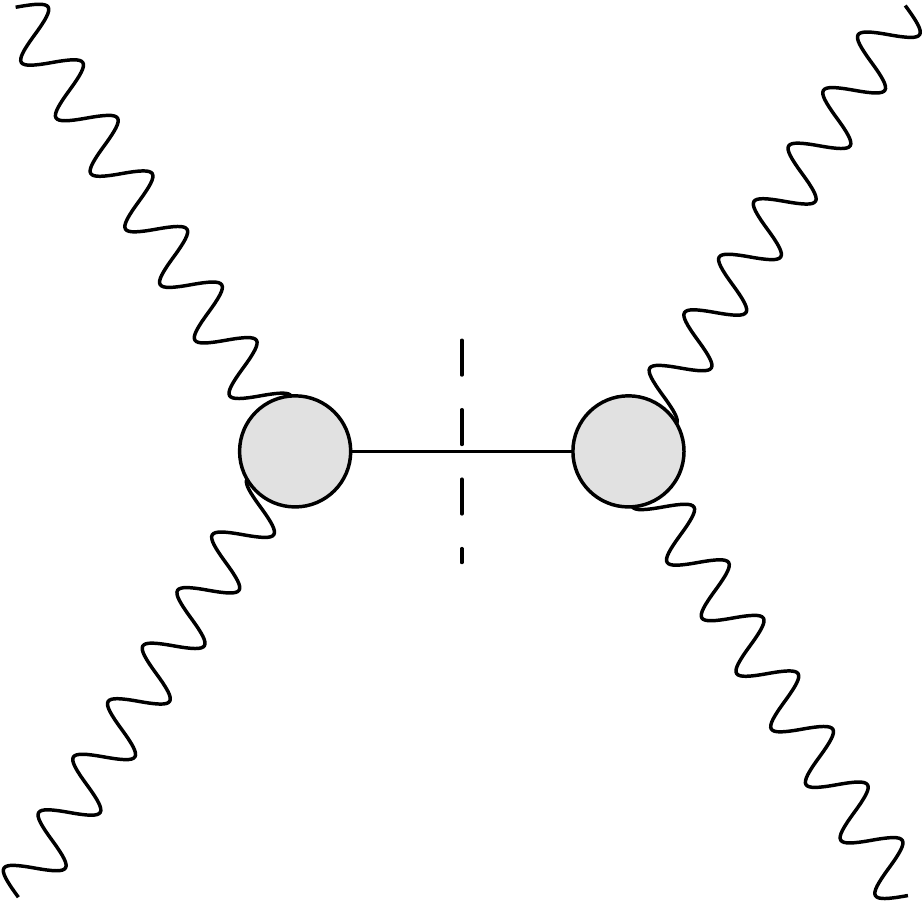}
		 \quad + \quad
		\includegraphics[width=2.5cm,valign=c]{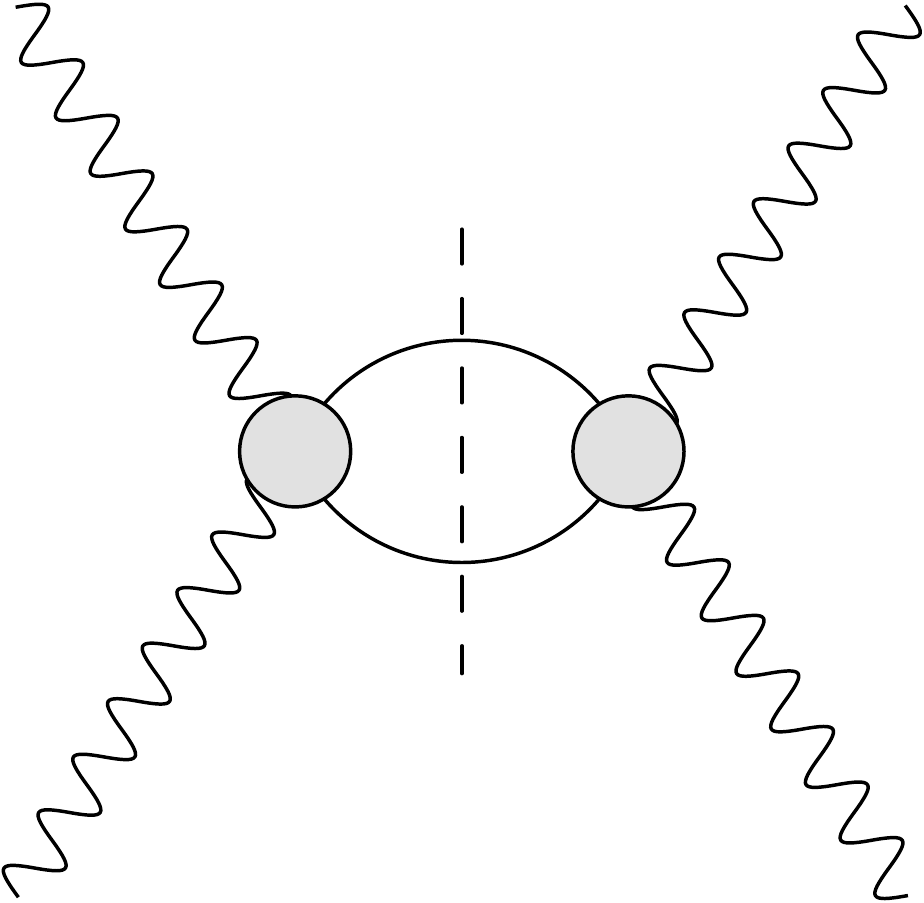}
		\quad + \quad \ldots
	\end{align*}
	\caption{Intermediate states in the direct channel: pion pole and two-pion cut.}
	\label{img:HLbLIntermediateStates}
\end{figure}

\begin{figure}[t]
	\centering
	\begin{align*}
		\includegraphics[width=2.5cm,valign=c]{TwoParticleCut}
		 =
		\includegraphics[width=2.5cm,valign=c]{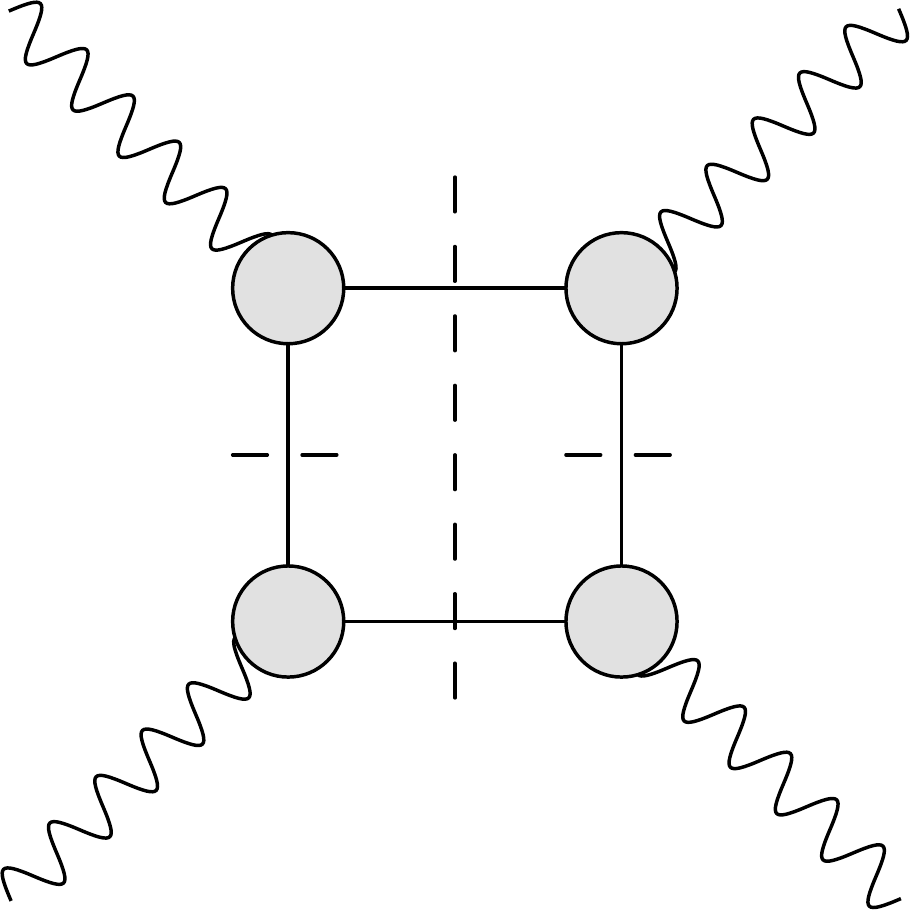}
		 +
		\includegraphics[width=2.5cm,valign=c,]{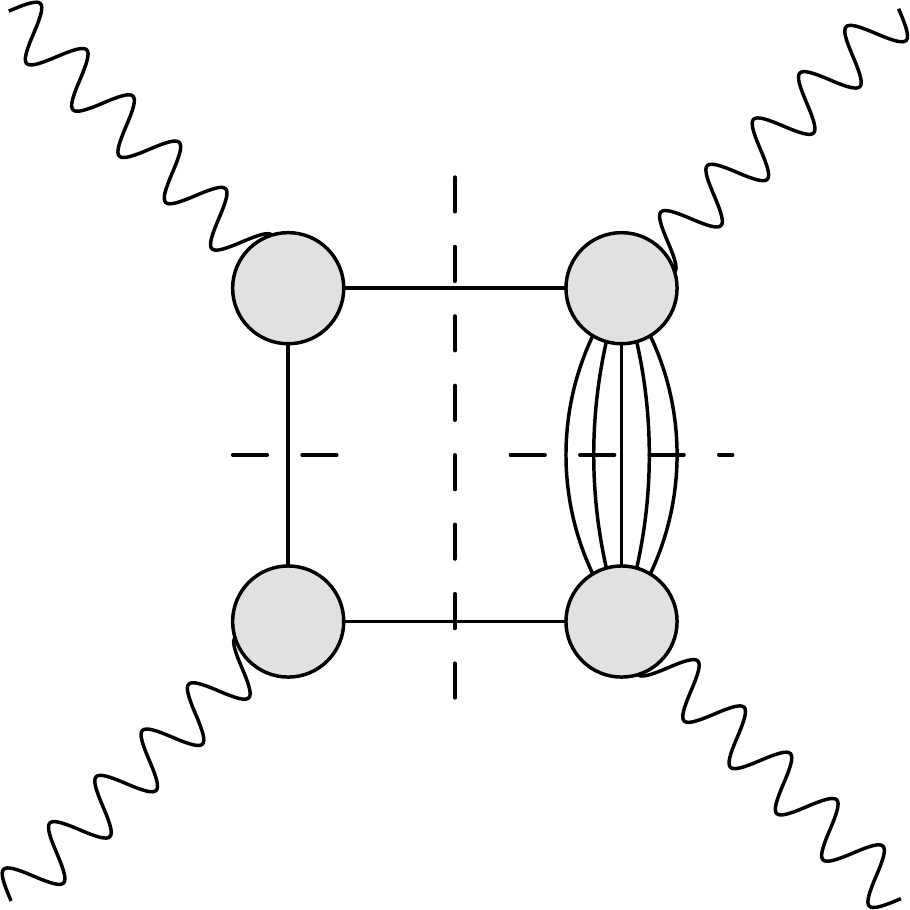}
		 +
		\includegraphics[width=2.5cm,valign=c]{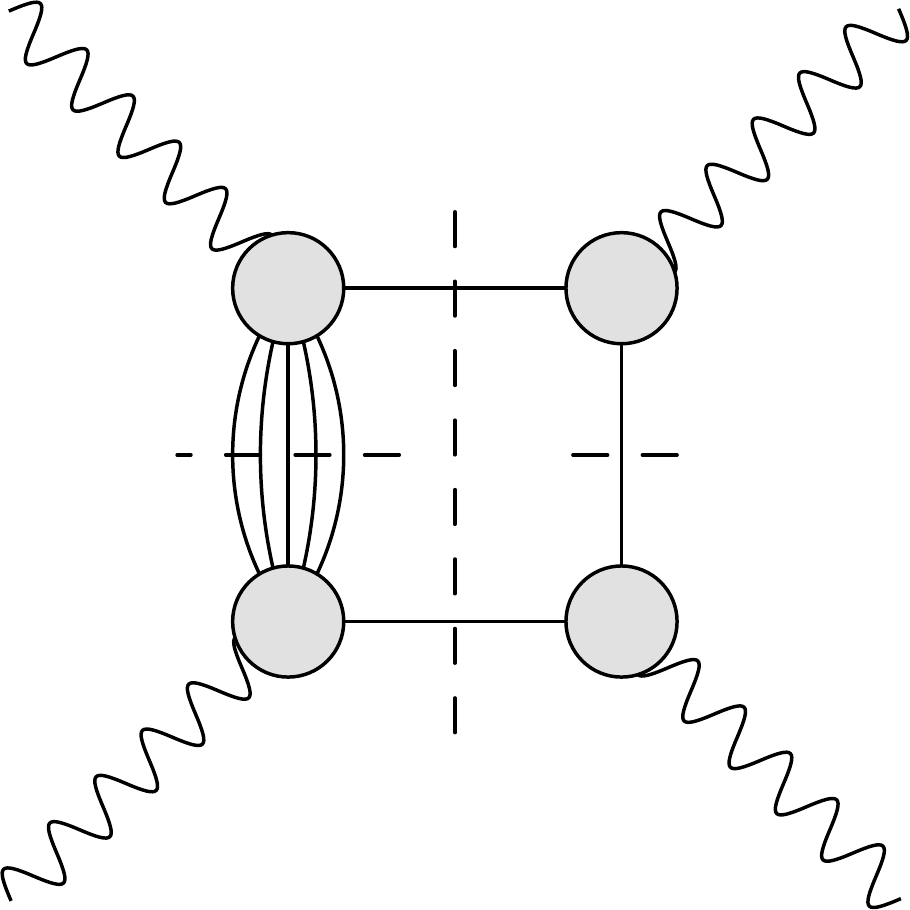}
		 +
		\includegraphics[width=2.5cm,valign=c]{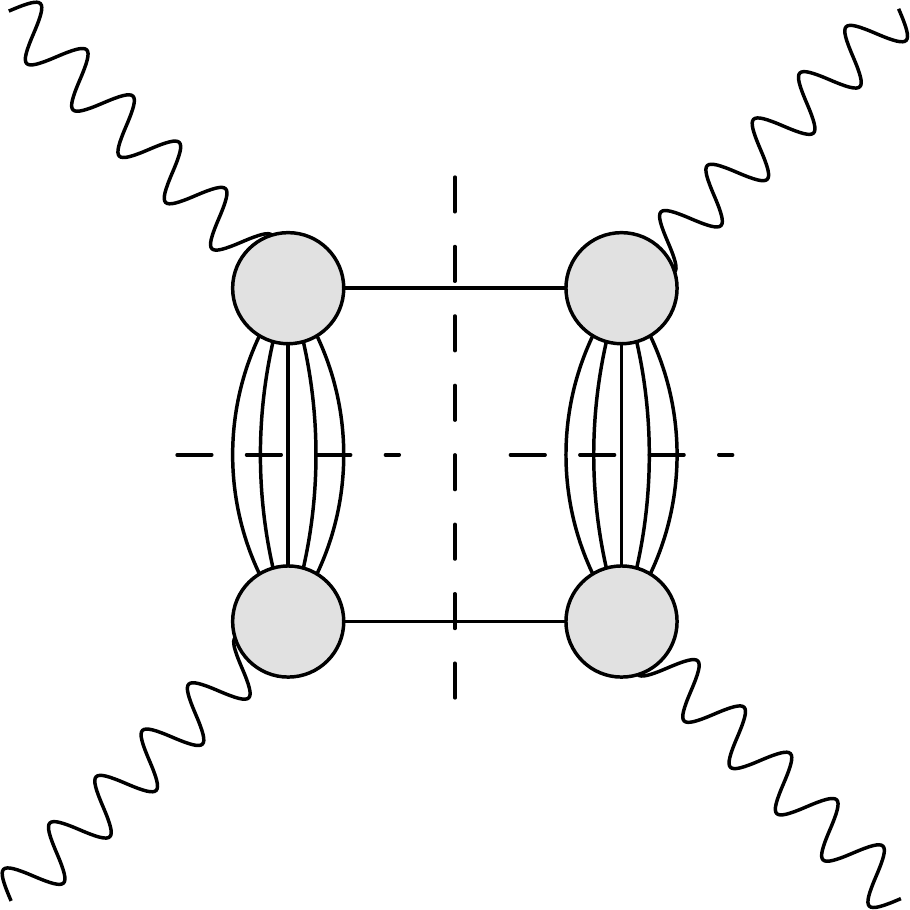}
		+ \ldots
	\end{align*}
	\caption{Two-pion contributions to HLbL. Further crossed diagrams are not shown explicitly.}
	\label{img:HLbLTwoPionContributions}
\end{figure}

Since it is impossible to include all possible intermediate states, the
question arises whether one can find an ordering principle for these, such
that one could concentrate on the most important ones, treat these
explicitly and neglect the rest. This is the approach which has been
adopted from the very start in this series of papers and has been described
first in~\cite{Colangelo:2014dfa}. The basis of this approach is not an
algebraically defined counting scheme, which so far has not been possible
to derive, but simply the observation that in all model calculations, the
importance of the contribution of an intermediate state decreases as the
corresponding threshold increases. As it is well known, the pion-pole
contribution is the dominating one overall and is more important than that
of other single-particle intermediate states, like the $\eta$ or the
$\eta'$ (with the former more important than the latter). The one-pion
contribution is more important than the two-pion contribution, which in
turn is more important than the two-kaon one, and so on. In order to set up
the dispersive calculation and obtain the bulk of the total contribution it
was argued in~\cite{Colangelo:2014dfa} that one could take into account
only one- and two-pion intermediate states, as illustrated in
Fig.~\ref{img:HLbLIntermediateStates}.
This allows one to break down the HLbL
contribution as follows
\begin{equation}
\Pi_{\mu\nu\lambda\sigma}(s,t,u)=\Pi_{\mu\nu\lambda\sigma}^{\pi^0\text{-pole}}(s,t,u)
+\Pi_{\mu\nu\lambda\sigma}^{\pi\text{-box}}(s,t,u) +
\Pi^{\pi \pi}_{\mu\nu\lambda\sigma}(s,t,u) + \cdots 
\end{equation}
where the first term $\Pi_{\mu\nu\lambda\sigma}^{\pi^0\text{-pole}}$ is the
one generated by the exchange of a $\pi^0$ in one of the channels ($s$ or
$t$ or $u$), the second one $\Pi_{\mu\nu\lambda\sigma}^{\pi\text{-box}}$
has two-pion discontinuities simultaneously in two channels ($s$ and $t$ or
$t$ and $u$ or $s$ and $u$), (see first diagram after the equal sign in
Fig.~\ref{img:HLbLTwoPionContributions}) whereas the third one
$\Pi^{\pi \pi}_{\mu\nu\lambda\sigma}$ has a two-pion cut only in one of the three
channels (second to fourth diagram after the equal sign in
Fig.~\ref{img:HLbLTwoPionContributions}). The ellipsis stands for
singularities with higher masses or thresholds. Note that even though we
restrict ourselves to pions, what really matters for the formalism is the
number of particles: applying the formalism to $\eta$ or $\eta'$ exchange,
or two-kaon intermediate states is a trivial extension. 

\subsection{Pion pole}\label{pionpole}
In a dispersive calculation the key step is to identify the singularities:
what defines the pion pole contribution is that the imaginary part is given
by a $\delta$-function, with a tensorial structure and strength fully
determined by the $\pi^0-\gamma \gamma$ vertex. This translates into the
following expression for the imaginary part in the $s$ channel:
\begin{align}
	\begin{split}
		\Im_s^\pi &\left( e^4 (2\pi)^4 \delta^{(4)}(q_1 + q_2 + q_3 - q_4) H_{\lambda_1\lambda_2,\lambda_3\lambda_4} \right) \\
			&= \frac{1}{2} \int \widetilde{dp} \; \< \gamma^*(-q_3,\lambda_3) \gamma^*(q_4,\lambda_4) | \pi^0(p) \>^* \< \gamma^*(q_1,\lambda_1) \gamma^*(q_2,\lambda_2) | \pi^0(p) \> 
	\end{split}
\end{align}
which, after reducing the matrix elements and using the definition of the
pion transition form factor 
\begin{align}
	i \int d^4x \; e^{i q x} \< 0 | T \{ j^\mu_\mathrm{em}(x) j^\nu_\mathrm{em}(0) \} | \pi^0(p) \> = \epsilon^{\mu\nu\alpha\beta} q_\alpha p_\beta \mathcal{F}_{\pi^0\gamma^*\gamma^*}\big(q^2, (q-p)^2\big) ,
\end{align}
leads to
\begin{align}
	\begin{split}
		\Im_s^\pi \Pi^{\mu\nu\lambda\sigma} &= - \frac{1}{2} \begin{aligned}[t]
				& \int \widetilde{dp} \; (2\pi)^4
                  \delta^{(4)}(q_1+q_2-p) \epsilon^{\mu\nu\alpha\beta}
                  \epsilon^{\lambda\sigma\gamma\delta} {q_1}_\alpha
                          {q_2}_\beta {q_3}_\gamma {q_4}_\delta
                          \mathcal{F}_{\pi^0\gamma^*\gamma^*}\big(q_1^2,q_2^2\big)
                          \mathcal{F}_{\pi^0\gamma^*\gamma^*}\big(q_3^2,q_4^2\big) \end{aligned}
                \\ 
			&= - \pi \delta( s - M_\pi^2 ) \epsilon^{\mu\nu\alpha\beta} \epsilon^{\lambda\sigma\gamma\delta} {q_1}_\alpha {q_2}_\beta {q_3}_\gamma {q_4}_\delta \mathcal{F}_{\pi^0\gamma^*\gamma^*}\big(q_1^2,q_2^2\big) \mathcal{F}_{\pi^0\gamma^*\gamma^*}\big(q_3^2,q_4^2\big) .
	\end{split}
\end{align}
Finally we only need to project this expression onto the scalar functions
$\Pi_i$, which leads to
\begin{align}
	\rho_{i;s}^t &= \left\{ \begin{matrix} \mathcal{F}_{\pi^0\gamma^*\gamma^*}\big(q_1^2,q_2^2\big) \mathcal{F}_{\pi^0\gamma^*\gamma^*}\big(q_3^2,q_4^2\big) & i = 1 , \\
									0 & i \neq 1 , \end{matrix} \right.
\end{align}
and, analogously,
\begin{align}
	\rho_{i;u}^t &= \left\{ \begin{matrix} \mathcal{F}_{\pi^0\gamma^*\gamma^*}\big(q_1^2,q_4^2\big) \mathcal{F}_{\pi^0\gamma^*\gamma^*}\big(q_2^2,q_3^2\big) & i = 3 , \\
									0 & i \neq 3  \end{matrix} \right.
\end{align}
for the two contributions proportional to a $\delta$-function in the
fixed-$t$ dispersion relation (as indicated by the superscript). By
considering also fixed-$s$ and fixed-$u$ dispersion relations and
symmetrizing the result, we obtain the following expression for the total
pion-pole contribution:
\begin{equation}
\Pi_i^{\pi^0\text{-pole}}(s,t,u) =
  \frac{\rho_{i,s}}{s-M_\pi^2} + \frac{\rho_{i,t}}{t-M_\pi^2} +
  \frac{\rho_{i,u}}{u-M_\pi^2}
\end{equation}
where
  \begin{align}
	\begin{split}
		\rho_{i,s} &= \delta_{i1} \; \mathcal{F}_{\pi^0\gamma^*\gamma^*}\big(q_1^2,q_2^2\big) \mathcal{F}_{\pi^0\gamma^*\gamma^*}\big(q_3^2,q_4^2\big) , \\
		\rho_{i,t} &= \delta_{i2} \;
                \mathcal{F}_{\pi^0\gamma^*\gamma^*}\big(q_1^2,q_3^2\big)
                \mathcal{F}_{\pi^0\gamma^*\gamma^*}\big(q_2^2,q_4^2\big) ,
                \\
		\rho_{i,u} &= \delta_{i3} \; \mathcal{F}_{\pi^0\gamma^*\gamma^*}\big(q_1^2,q_4^2\big) \mathcal{F}_{\pi^0\gamma^*\gamma^*}\big(q_2^2,q_3^2\big) . \\
	\end{split}
\nonumber
\end{align}
Only the first three of the 54 scalar functions receive a contribution from
the pion pole. 

This result can now be inserted into our master formula
Eq.~(\ref{eq:MasterFormula3Dim}) which provides a fully explicit
representation of the HLbL pion-pole contribution to $(g-2)_\mu$ as a
three-dimensional integral with as hadronic matrix element in the integrand
the pion transition form factor. Such a representation is not new and was
first derived in Ref.~\cite{Knecht:2001qf}, but the fact that our result
agrees with it represents a welcome check on our master formula which is
much more general than that. 

We will not dwell on the numerics here, because we have nothing new to
contribute or to report: the main difficulty in improving the
numerical evaluation of this contribution is related to obtaining a
reliable representation of the pion transition form factor for both photons
off-shell. Efforts in this direction are being made both with a dispersive
approach~\cite{Hoferichter:2014vra} (for related work providing essential
input see~\cite{Hoferichter:2012pm,Hoferichter:2017ftn}) as well as on the
lattice~\cite{Gerardin:2016cqj,GerardinLat17}, as already mentioned in
Sect.~\ref{sec:latticeMainz}.

\section{Pion Box}\label{pi-box}
\begin{figure}
\centering
\includegraphics[width=0.2\linewidth]{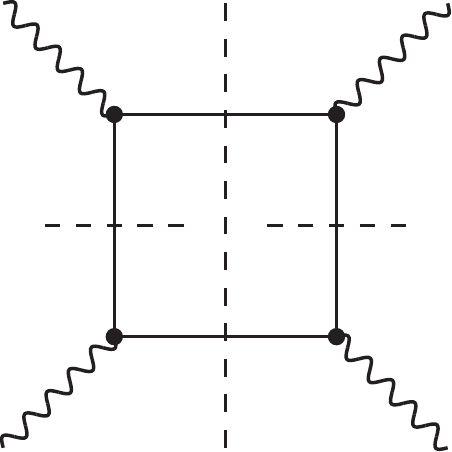} \quad
\includegraphics[width=0.2\linewidth]{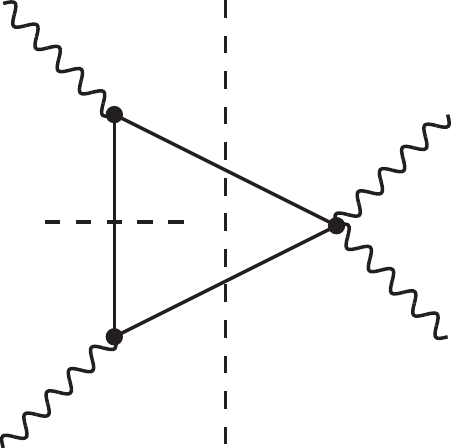}  \quad
\raisebox{0.77cm}{\includegraphics[width=0.2\linewidth]{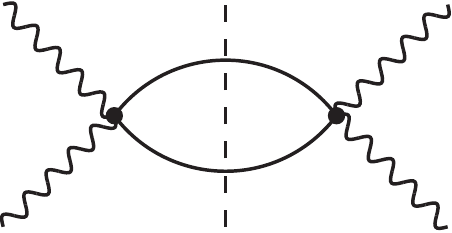}}
\caption{Scalar QED diagrams contributing to light-by-light scattering to
  one loop.}
\label{fig:sQED}
\end{figure}

The second contribution to the HLbL tensor and to $(g-2)_\mu$ which we
consider is the one given by the so-called ``pion box''. By this we mean a
contribution generated by a simultaneous cut due to two-pion intermediate
states in two of the three Mandelstam channels. Since the singularity is
completely determined by the configuration in which the intermediate states
are on-shell, and since in this case, as illustrated in
Fig.~\ref{img:HLbLTwoPionContributions}, all four pions in the diagram
contribute to the singularity and have to be put on-shell, the only unknown
hadronic matrix element in this contribution is the matrix element of an
electromagnetic current between two on-shell pions, which is (if we neglect
isospin breaking) nothing but the vector form factor of the pion:
\begin{equation}
\langle \pi^i(p_2)| \bar q \lambda_3 \gamma_\mu q | \pi^j(p_1) \rangle = i
\varepsilon^{i3j} (p_1+p_2)_\mu F^V_\pi((p_1-p_2)^2) \; \; .
\end{equation}
Since the $q_i^2$ variables are completely independent of the (two
independent) Mandelstam variables, and the singularities depend only on the
latter, the $q_i^2$ dependence does not play a role when one
reconstructs the full pion-box contribution in terms of its singularity.
Since the latter singularity is completely identical to the one
appearing in the scalar-QED (sQED) one-loop contribution to HLbL, we can
express the contribution of the pion box as follows:
\begin{equation}
\Pi_{\mu\nu\lambda\sigma}^{\pi\text{-box}}(s,t,u)=F^V_\pi(q_1^2)F^V_\pi(q_2^2)F^V_\pi(q_3^2)
F^V_\pi(q_4^2) \Pi_{\mu\nu\lambda\sigma}^{\text{sQED}}(s,t,u) \; \; .
\label{eq:pion-box}
\end{equation}
The statement that the singularities of the pion box are identical to those
of the one-loop sQED calculation may appear puzzling at first, especially
if one considers the Feynman diagrams in sQED, which are shown in
Fig.~\ref{fig:sQED}, but has been proven explicitly
in~\cite{Colangelo:2015ama}. There the sQED one-loop contribution was
calculated explicitly and then projected onto the BTT basis functions: the
corresponding Mandelstam representation obtained for the scalar functions
showed only singularities of the box type. This means that the seagull
vertex is only there to restore gauge invariance, and that if one works
with a gauge-invariant set of structures (like the BTT one), there is no
remnant of the seagull vertex in the singularity structure of the
corresponding scalar function. 
An even simpler case where a similar phenomenon is also visible, and
probably in a clearer way, is the
$\gamma^* \gamma^* \to \pi \pi$ process. As discussed in Sect. 2.6. of
Ref.~\cite{Colangelo:2015ama}, when one projects the sQED tree-level
contributions to this process, the corresponding scalar functions only have
pole singularities and no nonsingular term, despite what one would be led
to think by looking at the seagull vertex. Here also the latter only serves
the purpose of restoring gauge invariance at the Feynman-diagram level, and
when one works with a gauge-invariant basis there is no trace of it anymore.

Projecting representation~(\ref{eq:pion-box}) onto the BTT basis and taking
the limit of $(g-2)_\mu$ kinematics we obtain
a very practical representation of the pion box in terms of two-dimensional
Feynman-parameter integrals
\begin{align}
	\bar \Pi_i^{\pi\text{-box}}(q_1^2,q_2^2,q_3^2) = F_\pi^V(q_1^2) F_\pi^V(q_2^2) F_\pi^V(q_3^2) \frac{1}{16\pi^2} \int_0^1 dx \int_0^{1-x} dy  I_i(x,y) ,
\end{align}
where the integrands $I_i(x,y)$ have very compact expressions which are
given explicitly in appendix C of Ref.~\cite{Colangelo:2017fiz}. Inserting
these expressions in the master formula, Eq.~(\ref{eq:MasterFormula3Dim}),
we obtain
\begin{align*}
		a_\mu^{\pi\text{-box}}&= \frac{2 \alpha^3}{3 \pi^2} \int_0^\infty dQ_1 \int_0^\infty dQ_2 \int_{-1}^1 d\tau \sqrt{1-\tau^2} Q_1^3 Q_2^3 \begin{aligned}[t]
			& F_\pi^V(-Q_1^2) F_\pi^V(-Q_2^2) F_\pi^V(-Q_3^2) \\
			& \times \sum_{i=1}^{12} T_i(Q_1,Q_2,\tau) \bar \Pi_i^\mathrm{sQED}(Q_1,Q_2,\tau) . \end{aligned} \mytag
\end{align*}

\begin{figure}[t]
\centering
\includegraphics[height=5cm]{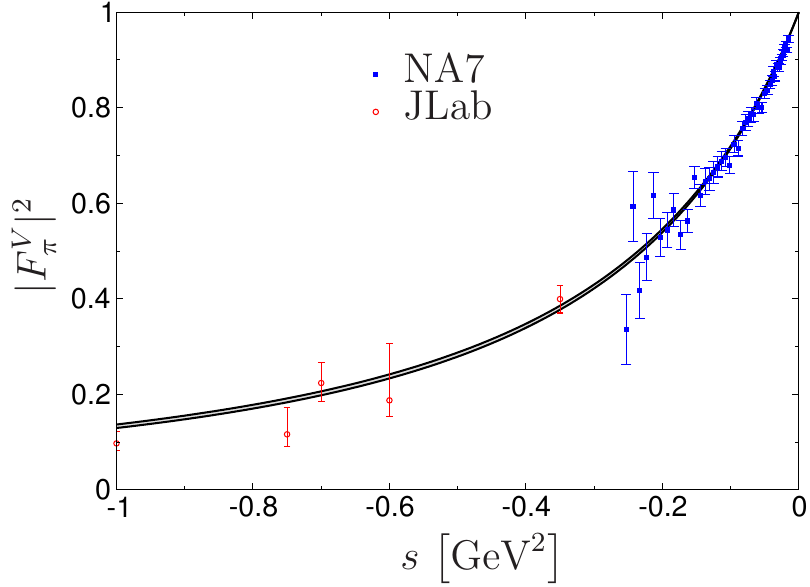}
\includegraphics[height=5cm]{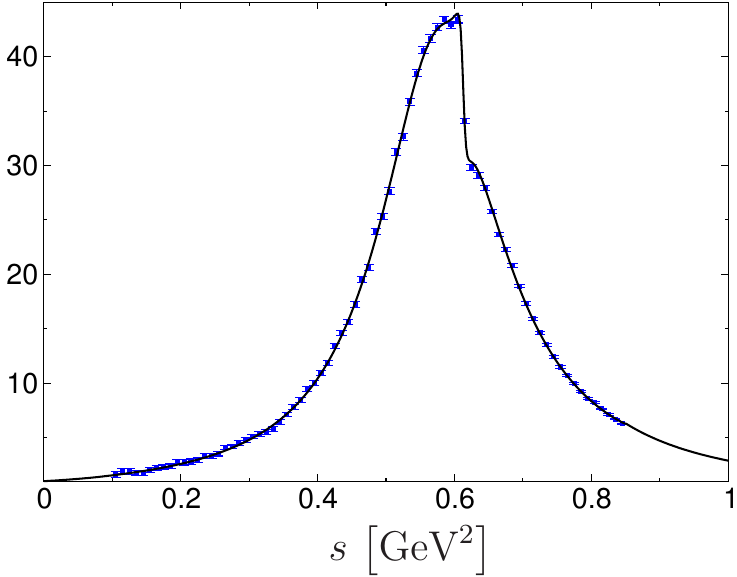}
\caption{Left: space-like pion form factor from our dispersive fit in
  comparison to data from NA7~\cite{Amendolia:1986wj} and
  JLab~\cite{Tadevosyan:2007yd,Blok:2008jy,Huber:2008id} (the latter are
  not included in the fit). The error band represents the variation
  observed between different time-like data sets. Right: pion form factor
  in the time-like region from the combined fit to NA7
  and~\cite{Ambrosino:2010bv}, 
chosen here for illustrative purposes only. Fits to the other time-like
data sets look very similar and lead to the same numerical results within
the accuracy quoted in~\eqref{pion_box}.} 
\label{fig:VFF}
\end{figure}

For a numerical evaluation one needs an explicit representation of the
vector form factor of the pion for spacelike momenta, and since about
$95\%$ of the final pion-box $(g-2)_\mu$ integral originate 
from virtualities below $1$ GeV, it is essential that the low-energy
properties be correctly reproduced. Experimentally, the available
constraints derive from $e^+e^-\to\pi^+\pi^-$ data, which determine the
time-like form
factor~\cite{Achasov:2006vp,Akhmetshin:2006bx,Aubert:2009ad,Ambrosino:2010bv,Babusci:2012rp,Ablikim:2015orh},
and space-like measurements by scattering pions off an electron
target~\cite{Dally:1982zk,Amendolia:1986wj}. We have also checked that our
representation is consistent with extractions of the space-like form factor
from $e^-p\to e^- \pi^+ n$
data~\cite{Horn:2006tm,Tadevosyan:2007yd,Blok:2008jy,Huber:2008id},
although due to the remaining model dependence of extrapolating to the pion
pole we do not use these data in our fits. To obtain a representation that
allows us to simultaneously fit space- and time-like data, and thereby
profit from the high-statistics form factor measurements motivated mainly
by the two-pion contribution to HVP, we adopt the formalism suggested
in~\cite{Leutwyler:2002hm,Colangelo:2003yw} (similar representations have
been used
in~\cite{DeTroconiz:2001rip,deTroconiz:2004yzs,Ananthanarayan:2013zua,Ananthanarayan:2016mns,Hoferichter:2016duk,Hanhart:2016pcd}). A
brief description of the formalism can be found in
Ref.~\cite{Colangelo:2017fiz}. Here we limit ourselves to a discussion of
the results, which are best illustrated by the two plots in
Fig.~\ref{fig:VFF}. 

The dispersive representation of the pion form factor is fixed by fitting
simultaneously the space-like data from~\cite{Amendolia:1986wj} as well
as one of the time-like data
sets~\cite{Achasov:2006vp,Akhmetshin:2006bx,Aubert:2009ad,Ambrosino:2010bv,Babusci:2012rp,Ablikim:2015orh}
(restricted to data points below $1$ GeV). All input parameters are varied
within reasonable bounds to check the dependence of the results on these.
We find that the results for the space-like form factor are extremely
stable against all these variations, the largest effect being produced by the
differences between the time-like data sets. For the accuracy required in
HLbL scattering we can therefore simply take the largest variation among them as
an uncertainty estimate, without having to perform a careful investigation
of the statistical and systematic errors that are crucial when combining
the different data sets for HVP.  
The result for the space-like form factor is shown in Fig.~\ref{fig:VFF},
leading to a numerical evaluation for the pion box of 
\begin{equation}
\label{pion_box}
a_\mu^{\pi\text{-box}}=-15.9(2)\times 10^{-11}.
\end{equation}
While the central value is in the ballpark of what had been obtained in
Ref.~\cite{Bijnens:1995xf}, and much larger (in absolute value) than the
estimate provided in Ref.~\cite{Hayakawa:1995ps} which was based on the
hidden gauge model, the greatest progress in our evaluation is in the error
reduction. Indeed all previous estimates assigned an uncertainty estimate
close to 100\% to this contribution, essentially because the description of
the photon off-shell dependence was considered to be pure model
work. Within our dispersive approach we were able to prove that the photon
off-shell dependence is rigorously described by the vector form factor of
the pion, which is very well known experimentally. The uncertainty in
Eq.~(\ref{pion_box}) is similar to the one obtained in the HVP calculation,
because it is essentially the same very precise data which constrain both
contributions. Which means that at the level of precision needed for the
HLbL contribution to $(g-2)_\mu$, the pion box is known essentially
exactly. 

\section{Partial waves, pion rescattering contribution}\label{rescatt}
The last contribution which we need to consider is the one coming from
diagrams two, three and four after the equal sign in
Fig.~\ref{img:HLbLTwoPionContributions}. These have two-pion
discontinuities in one channel but discontinuities of higher mass in the
other: the latter will not be treated explicitly, according to our
approximation scheme, but will be projected onto partial waves. In order to
treat these contributions we therefore need a formalism for dealing with
two-pion discontinuities in partial waves for HLbL. The case of $S$ waves
has been dealt with in Ref.~\cite{Colangelo:2014dfa} and is relatively
simple, but going beyond that has turned out to be a formidable task, which
has been solved and discussed in detail in
Ref.~\cite{Colangelo:2017fiz}. We list below here in the form of bullet
points some of the key reasons why this is so complicated:
\begin{itemize}
\item unitarity relations are diagonal in a helicity amplitude basis and
  the relation between the BTT (redundant, 54 elements) set and the helicity
  basis is neither unique nor invertible;
\item the helicity basis relevant for $(g-2)_\mu$ is the one with one
  on-shell photon, which has only 27 elements (half of the BTT set);
\item
  in the limit $q_4^2, q_4^\sigma \to 0$ of the HLbL tensor the number of
  independent elements of the BTT set drops from 41 to 27;
\item there is freedom in the choice of this subset (which we call a
  singly-on-shell basis); 
\item
  the transformation from helicity amplitudes to the singly-on-shell
  basis is easy to derive, but what one needs is the inverse of that;
\item
  inverting this relation (a $27\times 27$ matrix) is a lot more
  complicated but was done (analytically) in~\cite{Colangelo:2017fiz}; 
\item
  the arbitrariness in the choice of the 27 elements of the
  singly-on-shell basis would seem to affect the final result at first
  sight, but in fact it does not, because of sum rules;
\item
  these sum rules follow from the assumption that the HLbL tensor
  has a uniform behavior at short distances.
\end{itemize}
Understanding all this has not been easy, and it has been very important---
even mandatory--- to put the formalism under test, and the case of the pion
box has been invaluable for that. We have projected the pion box onto
partial waves and checked numerically whether the resummation of the
partial waves (carried out only up to a finite number of course) approached
the total, which can be calculated in one go, as discussed above. The test
was passed and the formalism is now ready to be used.

As a first numerical application we considered only $S$ waves. Their
contribution to the scalar functions can be expressed as follows:
 \begin{align}
  \begin{split}
		\hat\Pi_4^{S} &=\! \frac{1}{\pi} \int_{4M_\pi^2}^\infty  ds' \frac{-2}{\lambda_{12}(s')(s'-q_3^2)^2} \Big( 4s' \Im h^0_{++,++}(s') - (s'+q_1^2-q_2^2)(s'-q_1^2+q_2^2) \Im h^0_{00,++}(s') \Big)  \\
\hat\Pi_5^{S} &=\! \frac{1}{\pi} \int_{4M_\pi^2}^\infty dt' \frac{-2}{\lambda_{13}(t')(t'-q_2^2)^2} \Big( 4t' \Im h^0_{++,++}(t') - (t'+q_1^2-q_3^2)(t'-q_1^2+q_3^2) \Im h^0_{00,++}(t') \Big)  \\
\hat\Pi_6^{S} &=\! \frac{1}{\pi} \int_{4M_\pi^2}^\infty du' \frac{-2}{\lambda_{23}(u')(u'-q_1^2)^2} \Big( 4u' \Im h^0_{++,++}(u') - (u'+q_2^2-q_3^2)(u'-q_2^2+q_3^2) \Im h^0_{00,++}(u') \Big)  \\
		\hat\Pi_{11}^{S} &=\! \frac{1}{\pi} \int_{4M_\pi^2}^\infty du' \frac{4}{\lambda_{23}(u')(u'-q_1^2)^2} \Big( 2 \Im h^0_{++,++}(u') - (u'-q_2^2-q_3^2) \Im h^0_{00,++}(u') \Big)  \\
\hat\Pi_{16}^{S} &=\! \frac{1}{\pi} \int_{4M_\pi^2}^\infty dt' \frac{4}{\lambda_{13}(t')(t'-q_2^2)^2} \Big( 2 \Im h^0_{++,++}(t') - (t'-q_1^2-q_3^2) \Im h^0_{00,++}(t') \Big)  \\
\hat\Pi_{17}^{S} &=\! \frac{1}{\pi} \int_{4M_\pi^2}^\infty ds' \frac{4}{\lambda_{12}(s')(s'-q_3^2)^2} \Big( 2 \Im h^0_{++,++}(s') - (s'-q_1^2-q_2^2) \Im h^0_{00,++}(s') \Big) 
	\end{split}
\end{align}
where $\Im h^0_{00,++}$ and $\Im h^0_{++,++}$ are the imaginary parts of the
$S$-wave helicity amplitudes (the subscripts indicate the helicities of the
four photons) of the HLbL scattering amplitude, which unitarity relates to
(products of) the $S$-wave helicity amplitudes of $\gamma^* \gamma^* \to  
\pi \pi$. Unfortunately, the latter amplitudes, though measurable in
principle, have not been measured yet. There are very good data for
on-shell photons and some for singly-on-shell, but none for both photons
off-shell. 

To obtain our first numerical estimate we therefore proceeded as follows:
we considered the Roy-Steiner equations for $\gamma^* \gamma^* \to \pi
\pi$~\cite{Colangelo:2014dfa}, which needs as input an explicit
representation of the left-hand cut (as well as the $\pi \pi$ phase shifts,
which are known,
however~\cite{Colangelo:2001df,Caprini:2011ky,GarciaMartin:2011cn}), for
arbitrary photon virtualities. If one only considers the contribution to
the left-hand cut coming from the pion pole, the photon $q^2$ dependence is
again completely given by the pion vector form factor. Extensions to other
contributions to the left-hand cut will require additional input and will
be considered later.

For the concrete numerical implementation we have used the simplified
representation of the $\pi\pi$ phase
shifts based on the modified inverse-amplitude
method~\cite{GomezNicola:2007qj}, for the main reason that it has a
  simple analytic expression which is convenient to use in combination with
  Muskhelishvili--Omn\`es methods. This reproduces at the same time
  the low-energy properties of the phase shifts as well as pole  
position and couplings of the $f_0(500)$ resonance to a good accuracy. This
phase shift departs from the correct one just below the $K\bar K$ threshold
because it does not feature the sharp rise due to the $f_0(980)$ resonance
but continues flat with a smooth high-energy behavior. A full-fledged 
evaluation of the $f_0(980)$ resonance would require a proper treatment of
the $K\bar K$ channel, which is beyond the scope of this first estimate.
To control the dependence on the high-energy input we have introduced a
cutoff in the integral and have varied it from 1 GeV up to infinity.

\begin{table}[t]
\renewcommand{\arraystretch}{1.3}
\centering
\begin{tabular}{crrrr}
\hline
cutoff & $1\GeV$ & $1.5\GeV$ & $2\GeV$ & $\infty$\\ 
\hline
$I=0$ & $-9.2$ & $-9.5$ & $-9.3$ & $-8.8$\\
$I=2$ & $2.0$ & $1.3$ & $1.1$ & $0.9$\\
sum & $-7.3$ & $-8.3$ & $-8.3$ & $-7.9$\\
\hline
\end{tabular}
\caption{$S$-wave rescattering corrections to $a_\mu^{\pi\text{-box}}$, in units of $10^{-11}$, for both isospin components and in total.}
\label{tab:rescatt}
\end{table}

The results for the rescattering contribution, summarized in
Table~\ref{tab:rescatt}, are stable over a wide range of cutoffs,
indicating that our input for the $\gamma^*\gamma^*\to\pi\pi$ partial waves
reliably unitarizes the Born-term left-hand cut (LHC), which should indeed
dominate at low energies. In addition, we checked that the only sum rule
that receives $S$-wave contributions is already saturated at better than
$90\%$, completely in line with the expectation that the sum rules will be
fulfilled only after partial-wave resummation. The isospin-$0$ part of the
result can be interpreted as a model-independent implementation of the
contribution from  
the $f_0(500)$ of about
$-9\times 10^{-11}$ to HLbL scattering in $(g-2)_\mu$. In total, we obtain
for the $\pi\pi$-rescattering effects related to the pion-pole LHC 
\begin{equation}
\label{amupipi}
a_{\mu,J=0}^{\pi\pi,\pi\text{-pole LHC}}=-8(1)\times 10^{-11},
\end{equation}
where the error is dominated by the uncertainties related to the asymptotic
parts of the integral. 

\section{Conclusions and outlook}\label{concl}
After a quick overlook at ongoing lattice calculations of the
HLbL contribution to the $(g-2)_\mu$ we have concentrated on the dispersive
approach and briefly summarized its basic steps. We have then described the
ingredients of the first numerical evaluation of the pion box and related
rescattering corrections in the $S$ wave. Adding the two contributions
together we have obtained 
\begin{equation}
a_\mu^{\pi\text{-box}} + a_{\mu,J=0}^{\pi\pi,\pi\text{-pole
    LHC}}=-24(1)\times 10^{-11}, 
\end{equation}
which represents the first precise and model-independent evaluation of
these two contributions. Compared to other sources of uncertainty in HLbL
the error in the estimate provided here is negligible. This is of course
only a first step and in order to complete the calculation based on the
dispersive approach, there are other contributions which will need to be
considered. The most important one is the one due to the pion-pole and for
that it is the pion transition form factor which represents the crucial
input quantity. Efforts to evaluate the latter on the basis of a dispersion
relation are ongoing~\cite{Hoferichter:2014vra}. But also for what concerns
the two-pion contribution there is still more work to be done, in
particular (i) by adopting a more realistic representation of the left-hand
cut, which in turn will require modeling the off-shell behavior of the
photons; (ii) including higher partial waves, in particular to $D$ wave,
which contains the very prominent $f_2(1270)$ resonance; (iii) even for the
$S$ wave, it is important to include a realistic description of the region
above 1 GeV, even if this will probably be subdominant; (iv) as mentioned
above doing this will only be possible if simultaneously considering the
contribution of two-kaon intermediate states; (v) finally, an estimate of 
contributions which go beyond two-pion intermediate states will be very
important in order to assess the robustness of the final numerical
evaluation. For example, three-pion intermediate states can be modeled by
means of axial resonance contributions.

These comments show that there is still a lot of work ahead of us before
being able to provide a complete estimate of the HLbL contribution to the
$(g-2)_\mu$ based on a dispersive approach. We are confident, however, that
this ambitious goal is now in sight, and that the availability of two
model-independent approaches (the lattice and the dispersive one) to the
calculation of this contribution is a very significant step ahead towards a
deeper understanding of the $(g-2)_\mu$ puzzle.

\section*{Acknowledgments}
It is a pleasure to thank the organizers for the invitation to a
very interesting Lattice conference and for their perfect
organization of the event. I gratefully acknowledge help in preparing the
slides on recent lattice progress in the calculation of HLbL by Christoph
Lehner and Hartmut Wittig, who have also read and given useful comments on the
manuscript. Participation to the conference and this work have been
supported by the Swiss National Science Foundation. M.H.\ is supported by
the DOE (Grant No.\ DE-FG02-00ER41132), M.P.\ by a Marie Curie
Intra-European Fellowship of the European Community's 7th Framework
Programme under contract number PIEF-GA-2013-622527, and P.S.\ by a grant
of the Swiss National Science Foundation (Project No.\ P300P2\_167751) and
of the DOE (Grant No. DE-SC0009919). 

\clearpage
\bibliography{Literature}
\end{document}